\documentclass[12pt]{article}
\usepackage{epsfig}
\usepackage{graphicx}
\def\Title#1{\begin{center} {\Large {\bf #1} } \end{center}}

\begin{document}

\Title{Pulse-to-pulse Orbit Jitter Propagation in Multi-bunch Operation at the KEK Accelerator Test Facility 2 (ATF2)}

\bigskip\bigskip


\begin{raggedright}  

{\it G.~B.~Christian}\\
John Adams Institute (JAI), Oxford, United Kingdom\\
{\it J.~Resta-L\'opez} \footnote{resta@ific.uv.es}\\
Instituto de F\'isica Corpuscular (IFIC), Valencia, Spain
\bigskip\bigskip
\end{raggedright}

\begin{abstract}
Pulse-to-pulse orbit jitter, if not controlled, can drastically degrade the luminosity in future linear colliders. The second goal of the ATF2 project at the KEK accelerator test facility is to stabilise the vertical beam position down to approximately 5$\%$ of the nominal rms vertical beam size at the virtual Interaction Point (IP). This will require control of the orbit to better than 1 micrometre at the entrance of the ATF2 final focus system. In this report simulation studies are presented for vertical jitter propagation through the ATF2 extraction line and final focus system, and the jitter is evaluated at the IP. For these studies pulse-to-pulse vertical jitter measurements using three stripline beam position monitors are used as initial inputs. These studies are performed for the case of a bunch-train with three bunches, but could easily be extended for a larger number of bunches. The cases with and without intra-train orbit feedback correction in the extraction line of ATF2 are compared.  
\end{abstract}

\section{Introduction}
The control of pulse-to-pulse orbit jitter for future linear colliders will be crucial to achieve the required design luminosity. Both beam position and angular jitter should be controlled along the beam delivery system during multi-bunch operation in order to stabilise the vertical beam position jitter to the nanometre level precision at the IP. 

\begin{figure}[hb]
   \centering
   \includegraphics*[width=14cm]{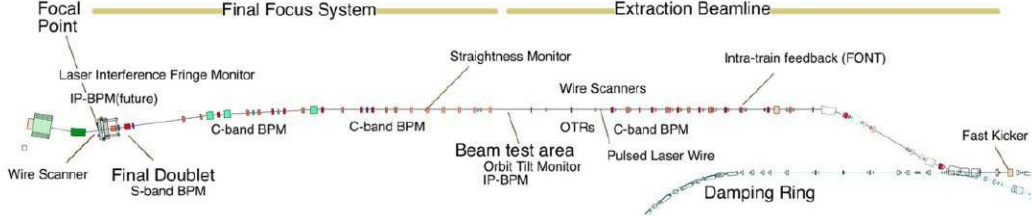}
   \caption{Layout of the ATF2 beamline, indicating the position of the major beam instrumentation systems.}
   \label{ATF2layout}
\end{figure}

The ATF2 final focus test beam facility \cite{ATF2} is currently progressing towards the achievement of transverse beam sizes of about 40~nm at the IP. At the same time, R\&D activities have also started to achieve the second ATF2 goal, i.e. the control of the beam position at the level of 5$\%$ of the rms vertical beam size at the IP. Figure~\ref{ATF2layout} shows a schematic of the ATF2 beamline. A two-phase intra-train Feedback (FB) system for position and angle correction has been installed in the extraction line of ATF2. This FB system is based on two kickers and three stripline Beam Position Monitors (BPMs), which allow the bunch-by-bunch measurement of $x$ and $y$ jitter in multi-bunch operation. 

Several beam tests of the intra-train FB system at the ATF2 were performed for three bunch trains during 2010. The vertical position jitter was measured for the cases with and without FB correction in the extraction line of ATF2. In this paper, using these jitter measurements as initial inputs, we compute the jitter propagation along the nominal ATF2 lattice, and predict the position and angle jitter at the IP.  


\section{Intra-train Feedback System}

\begin{figure}[t]
   \centering
   \includegraphics*[width=14cm]{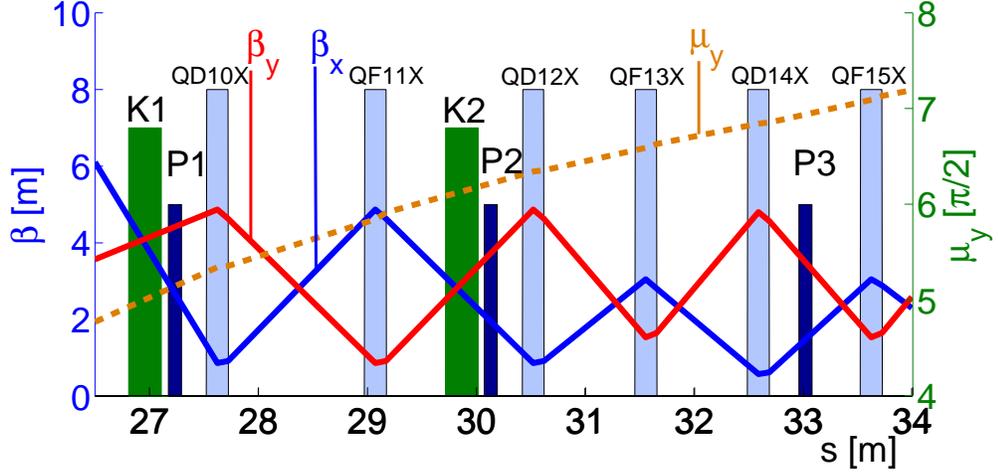}
   \caption{Schematic layout of the extraction line of the ATF2 beamline showing the relative locations of the FONT5 kickers (K1 and K2) and BPMs (P1, P2 and P3).}
   \label{FONTbeamline}
\end{figure}

In the context of the Feedback On Nano-second Timescales (FONT) project \cite{FONT1,FONT2, FONT3, FONT4, FONT5_1}, an ILC-like intra-train FB system prototype (FONT5) \cite{FONT5_1, FONT5_2} has been designed and tested in the extraction line of ATF2. A schematic of the FONT5 FB system elements in the ATF2 beamline is shown in Fig.~\ref{FONTbeamline}. The key components of this system are: a pair of stripline kickers (K1 and K2), located with $\pi/2$ phase advance in between them, for applying beam position and angle correction in the vertical phase space; three stripline BPMs for registering the beam orbit (P1, P2 and P3); and additional electronic components, such as FB circuits, fast amplifiers and data acquisition devices.  

The FONT5 system incorporates a digital feedback processor based on a state-of-the art Field Programmable Gate Array (FPGA) \cite{FPGA}. This allows the implementation of FB algorithms for simultaneous and coupled $y$ and $y'$ correction or, on the other hand, the configuration of two independent loops for $y$ and $y'$ separately. 


The FONT5 system has been tested at ATF2 to correct the incoming pulse-to-pulse jitter (jitter that is correlated between bunches) for 3-bunch trains.   

The three BPMs of the FONT system can be used to provide information of the transverse beam jitter along the ATF2 beamline. In the next section we try to answer the following questions: What is the jitter ratio between the cases with and without correction at any point downstream of the FONT region? Is the jitter reduction given by the FB system preserved at the virtual IP?       

\section{Vertical Jitter Measurements}

The FONT5 intra-train FB system was tested to correct the vertical position and angle jitter in the extraction line of ATF2. The ATF2 was operated to provide 1.3~GeV bunch-trains with 3 bunches, and bunch separation of 154~ns. The FB system was operated in coupled FB mode in order to correct simultaneously $y$ and $y'$, interleaving the measurements with FB switched off and on. The FB system measures the first bunch position and corrects the subsequent bunches. The figure of merit for this FB system is the reduction in the beam jitter, for correlated bunch-to-bunch jitter.

Position jitter measurements by the three FONT BPMs are shown in Table~\ref{measurements1}. This set of measurements for 1000 pulses corresponds to a test performed on the 16th April 2010. For this dataset the measured bunch-to-bunch jitter was particularly highly correlated. The incoming (feedback off) bunch-to-bunch correlation were measured to be $98\%$ for bunch 1--bunch 2 and $89\%$ for bunch 2--bunch 3.
 
The positions at P2 and P3 were used to calculate the angle distribution at P2, and from this the rms angle jitter at P2 (see Table~\ref{anglejitter}). The BPM resolution was estimated to be better than 0.4~$\mu$m for P2 and about 1~$\mu$m for P3. 
  
\begin{table}[hbt]
\begin{center}
   \caption{ Vertical beam jitter measurements by the FONT5 BPMs for each bunch in 3-bunch train operation. Data from 16th April 2010.}
   \begin{tabular}{lccc}
       \hline \hline
       Bunch $\#$ & BPM P1 & BPM P2 & BPM P3 \\
         {} & $\sigma_{(1)}$ [$\mu$m] & $\sigma_{(2)}$ [$\mu$m] & $\sigma_{(3)}$ [$\mu$m] \\         
       \hline
        1 (FB OFF/ON) & 3.3/3.4 & 2.4/2.2 & 3.4/3.2 \\
        2 (FB OFF/ON) & 3.2/3.3 & 2.3/0.4 & 3.3/1.8 \\
        3 (FB OFF/ON) & 3.3/3.5 & 2.5/1.1 & 3.3/1.6 \\      
       \hline
   \end{tabular}
   \label{measurements1}
\end{center}
\end{table}


\begin{table}[hbt]
   \begin{center}
   \caption{ Vertical angle beam jitter at BPM P2. Data from 16th April 2010.}
   \begin{tabular}{lc}
       \hline \hline
       Bunch $\#$ & BPM P2  \\
         {} & $\sigma'_{(2)}$ [$\mu$rad] \\         
       \hline
        1 (FB OFF/ON) & 1.9/1.7 \\
        2 (FB OFF/ON) & 1.9/0.65 \\
        3 (FB OFF/ON) & 1.9/0.72 \\      
       \hline
   \end{tabular}
   \label{anglejitter}
\end{center}
\end{table}

\section{Transverse Jitter Calculations}

Let us assume a beam pulse consisting of multiple bunches, each bunch centroid characterised by a vector $\mathbf{y}=(y,y')$ containing the information on vertical position and angle. If for each bunch an ensemble of $y$ and $y'$ measurements is performed over many beam pulses, the ensemble of a given bunch can be characterised statistically by the following covariance matrix: 

\begin{equation}
\Sigma=\langle(\mathbf{y}-\bar{\mathbf{y}})^T(\mathbf{y}-\bar{\mathbf{y}})\rangle=\left(\begin{array}{cc}
\langle y^2 \rangle & \langle yy' \rangle \\
\langle y'y \rangle & \langle y'^2 \rangle \\
\end{array} \right)\,\,,
\end{equation}

\noindent where $\mathbf{y}^T$ indicates the transpose of $\mathbf{y}$. We will consider normal distributions with zero mean value: $\bar{\mathbf{y}}=\mathbf{0}$.

The rms position and angle jitter of the bunch can be defined as $\sigma=\sqrt{\langle y^2 \rangle}$ and $\sigma'=\sqrt{\langle y'^2 \rangle}$, respectively.

The evolution of the covariance matrix between two points $s_1$ and $s_2$ of a transfer line is given by: 

\begin{equation}
\Sigma(s_2)=R\Sigma(s_1)R^T\,\,,
\label{covariancematrix}
\end{equation}

\noindent where $R$ is the transfer matrix from $s_1$ to $s_2$, and $R^T$ is the transpose of $R$. Here we consider only the 2-D transfer matrix in the vertical plane:

\begin{equation}
R=\left(\begin{array}{cc}
R_{33} & R_{34} \\
R_{43} & R_{44} \\
\end{array}\right)
\end{equation}

If $\Sigma$ is known at a certain position and the optical lattice is known, then the position and angular jitter can be evaluated at any other point of the lattice. 

Another way to calculate the jitter values along the beamline is using the covariance vector $(\langle y^2 \rangle, \langle y'^2 \rangle,\langle y y' \rangle)$ instead of the previous covariance matrix. If the optics of the transfer line is known, from the rms position measurements at three different BPMs, $\sigma_{(1)}=\sqrt{\langle y^2_1 \rangle}$, $\sigma_{(2)}=\sqrt{\langle y^2_2 \rangle}$ and $\sigma_{(3)}=\sqrt{\langle y^2_3 \rangle}$, the position and angular jitter can be calculated at any point of the beamline,

\begin{equation}
\left(\begin{array}{c}
\langle y^2_i \rangle \\
\langle y'^2_i \rangle \\
\langle y_iy'_i \rangle \\
\end{array} \right) = M^{-1} \left( \begin{array}{c}
\sigma^2_{(1)} \\
\sigma^2_{(2)} \\
\sigma^2_{(3)} \\
\end{array} \right)
\end{equation} 

\noindent where $M$ is the following matrix:

\begin{equation}
M = \left(\begin{array}{ccc}
R^2_{33,(1)} & R^2_{34,(1)} & 2R_{33,(1)}R_{34,(1)} \\
R^2_{33,(2)} & R^2_{34,(2)} & 2R_{33,(2)}R_{34,(2)} \\
R^2_{33,(3)} & R^2_{34,(3)} & 2R_{33,(3)}R_{34,(3)} \\
\end{array} \right) 
\end{equation}

\noindent where the terms $R_{33,(j)}$ and $R_{34,(j)}$ are the matrix terms corresponding to the transfer matrix from the $j$th BPM ($j=1,2,3$) to the point $i$ where we want to evaluate the jitter values: $\sigma_{(i)}=\sqrt{\langle y^2_i \rangle}$, $\sigma'_{(i)}=\sqrt{\langle y'^2_i \rangle}$. 

Knowing the rms position jitter at P2 and P3 and the rms angle jitter at P2 we can obtain the value of the covariance $\langle y y' \rangle$ at P2. In this way, the elements of the matrix $\Sigma$ are known at P2, and Eq.~(\ref{covariancematrix}) can be used to evaluate the value of the position and angle jitter along the ATF2 beamline. For this calculation we have obtained linear transfer matrices $R$ for the nominal optics using the code MAD \cite{MAD}. The nominal ATF2 optics in MAD format, version 4.3a, can be found in the web repository \cite{ATFrepository}. 

\begin{figure}[htb]
   \centering
   \includegraphics*[width=12cm]{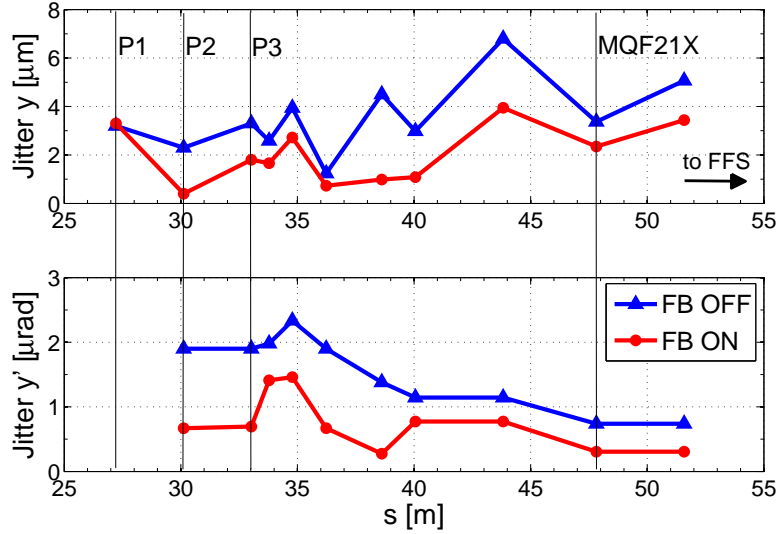}
   \caption{ Position (Top) and angle (Bottom) jitter propagation for bunch 2 along the ATF extraction line predicted from the rms jitter position measured by the FONT BPMs. The cases FB ON and OFF are compared.}
   \label{jitterpropagaEXT}
\end{figure}

\begin{figure}[htb]
   \centering
   \includegraphics*[width=13cm]{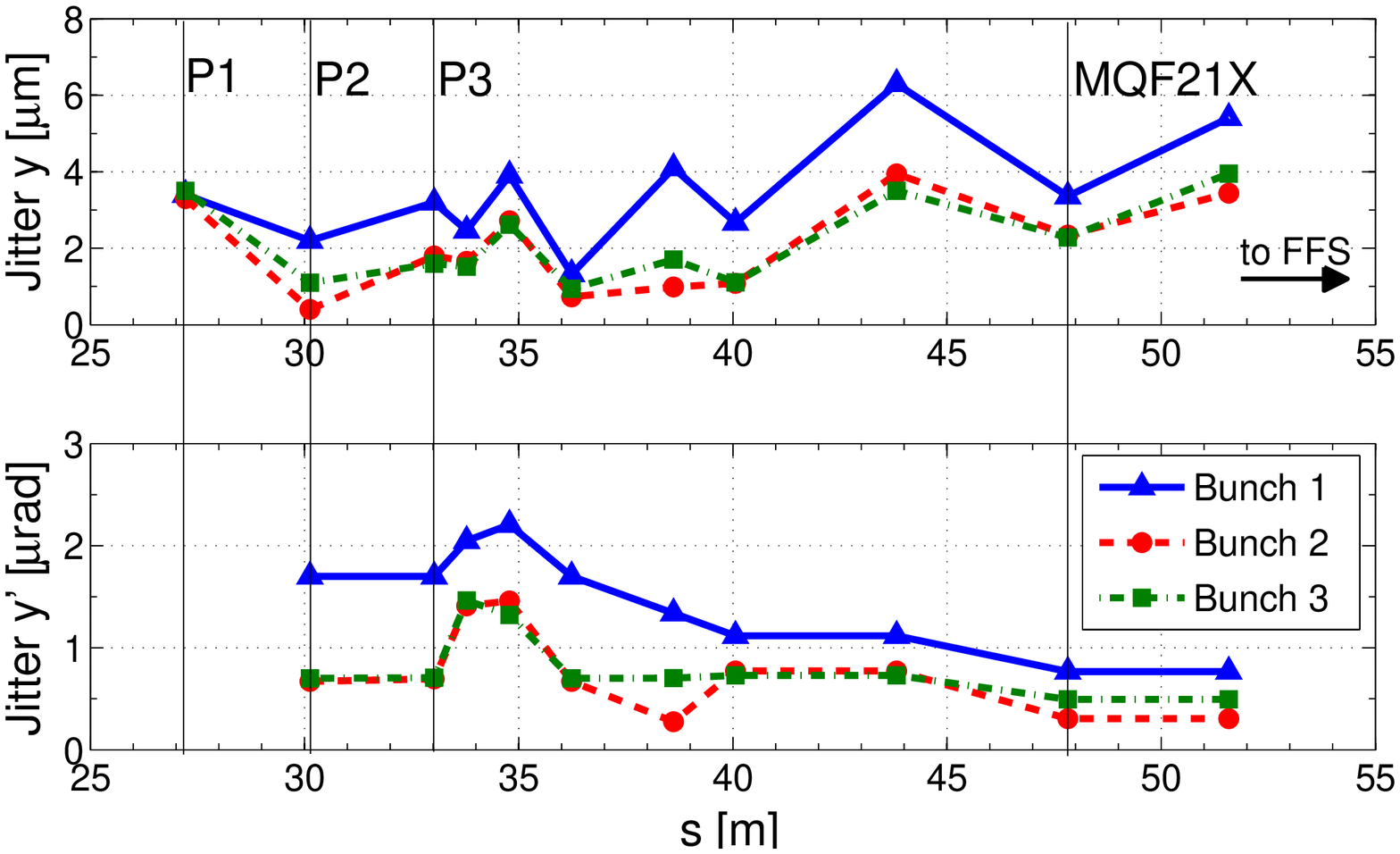}
   \caption{ Position (Top) and angle (Bottom) jitter propagation for bunch 1, 2 and 3 along the ATF extraction line predicted from the rms jitter position measured by the FONT BPMs, for the case with FB ON.}
   \label{jitterpropagabunches}
\end{figure}

Figure~\ref{jitterpropagaEXT} shows the prediction of the vertical position and angle jitter  for bunch 2 in the ATF2 extraction line downstream of the FONT region for the nominal lattice. No extra source of jitter has been assumed downstream of the FB system. In a similar way, Fig.~\ref{jitterpropagabunches} shows vertical jitters for each of the three bunches in the ATF2 extraction line, using the nominal lattice and for the case with FB switched on. According this results, the FB system reduces the vertical jitter of bunches 2 and 3, after measuring the jitter of bunch 1, and this jitter reduction remains along all the extraction line.     

\section{Tracking Simulations}

In the previous section, only the linear transport was taking into account for the jitter evaluation. For a more precise prediction of the vertical jitter at the virtual IP, tracking simulations of vertical offset distributions through the ATF2 beamline have been performed. These tracking simulations consider the nonlinearities of the beamline optics. This is important for the beam transport through the final focus system, where five sextupoles have been placed for local chromaticity correction.  

For the initial $y$ and $y'$ offset distribution in these tracking simulation studies, the following bivariate normal distribution has been generated at P2:

\begin{eqnarray}
y & = & [r_1]\sigma_{(2)}\,\,,\\
y'& = & [r_1]\sigma'_{(2)}\rho + [r_2]\sigma'_{(2)}\sqrt{1-\rho^2}\,\,,
\end{eqnarray} 

\noindent where $[r_1]$ and $[r_2]$ are standard normal random variables. $\sigma_{(2)}$ and $\sigma'_{(2)}$ are the rms position and angle jitter measured at P2, respectively. $\rho=\langle y y' \rangle/(\sigma_{(2)}\sigma'_{(2)})$ is the $y$-$y'$ correlation factor. For the case FB OFF  $\rho=0.85$, and for FB ON $\rho=0.03$.

For this tracking study the code MAD has been used. The simulation assumes no extra source of jitter downstream of the FB system. 

\begin{figure}[htb]
   \centering
   \includegraphics*[width=12cm]{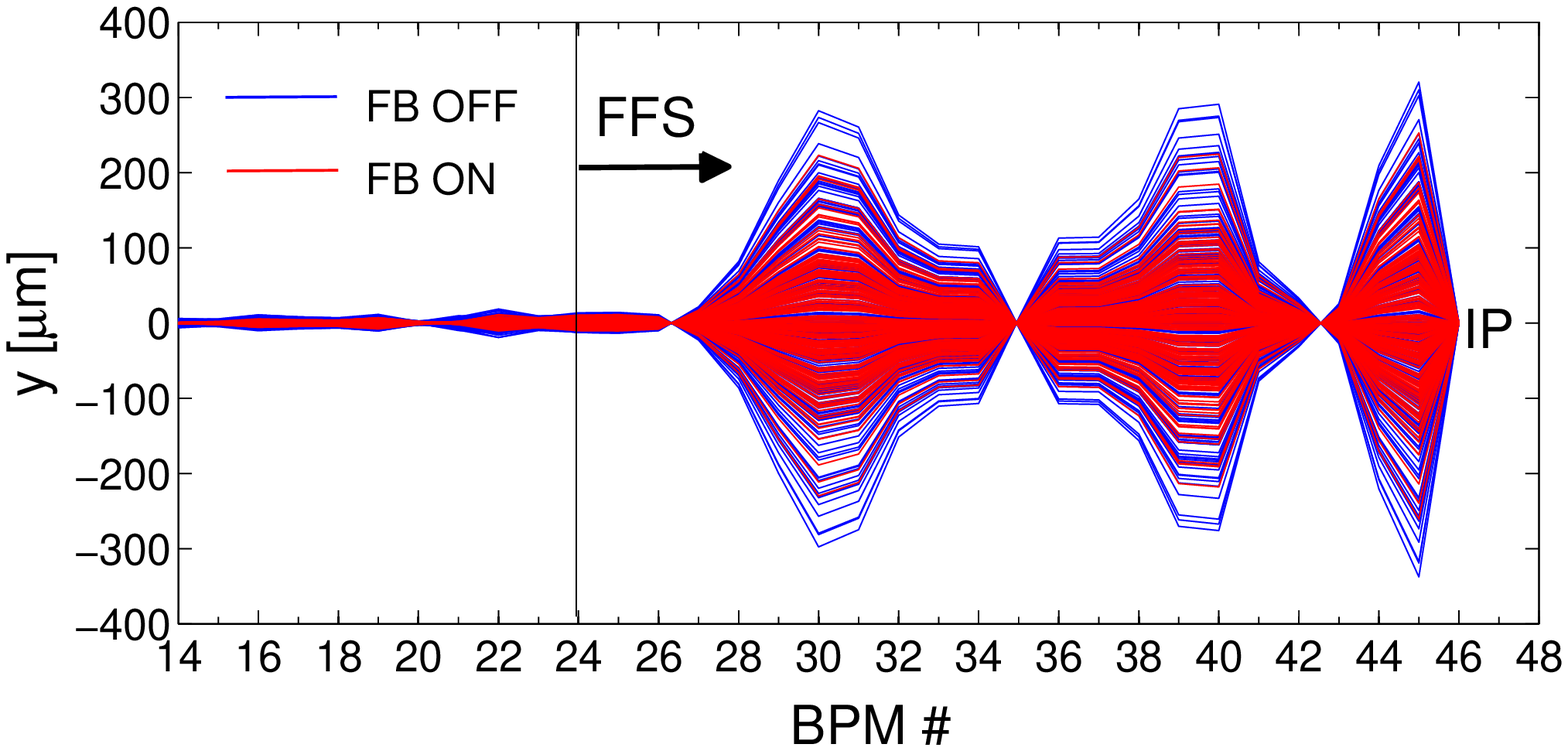}
   \includegraphics*[width=12cm]{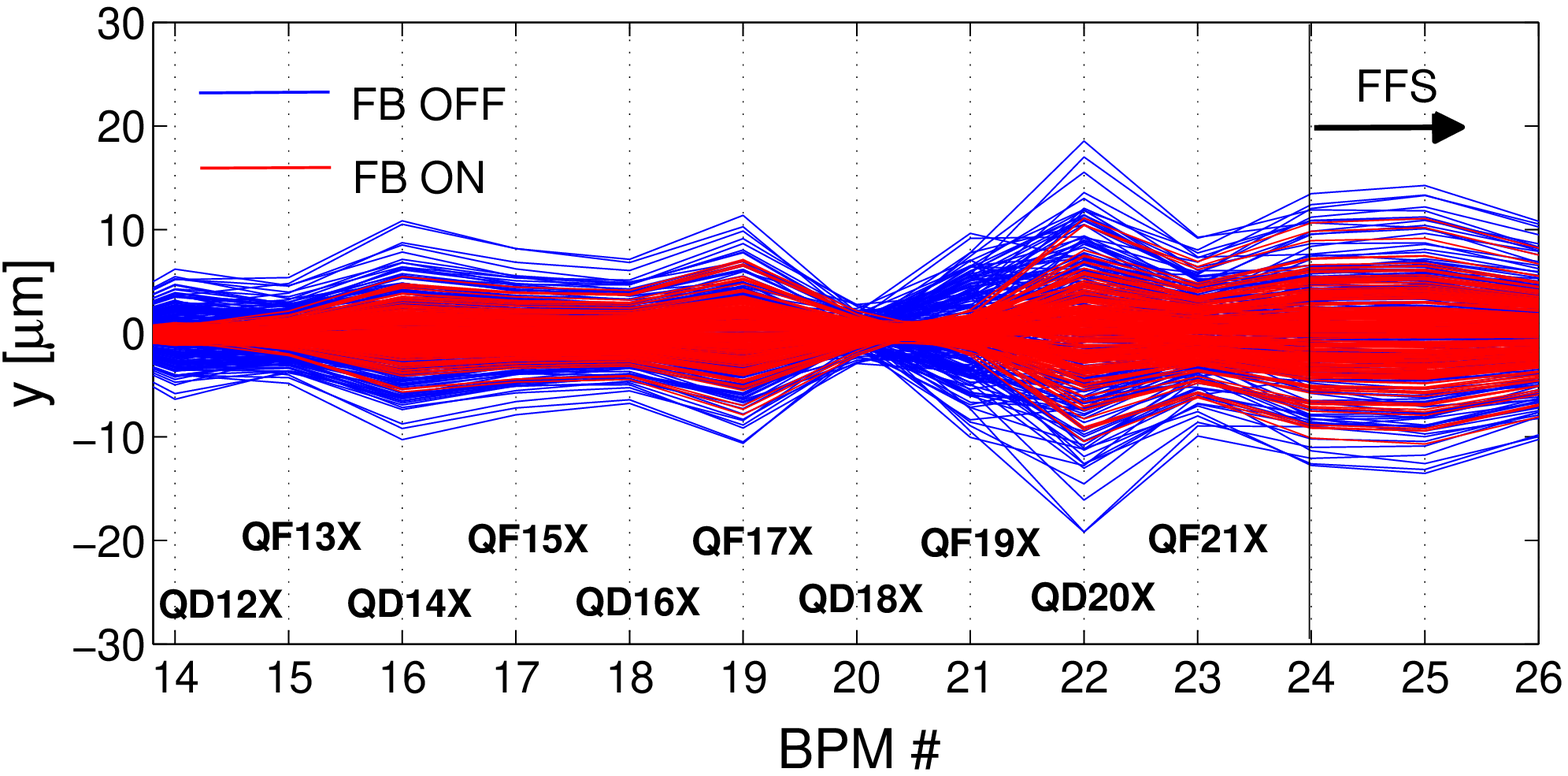}
   \caption{ Propagation of a normal distribution of 200 vertical offsets through the ATF2 beamline from the BPM P2 to the IP for the cases with FB OFF and FB ON (Top). Zoom of the extraction region (Bottom).}
   \label{Offsettracking}
\end{figure}

Figure~\ref{Offsettracking} illustrates the evolution of a distribution of 200 vertical offsets through the ATF2 beamline for bunch 2. A clear improvement is observed with FB ON. 

\begin{figure}[htb]
   \centering
   \includegraphics*[width=14cm]{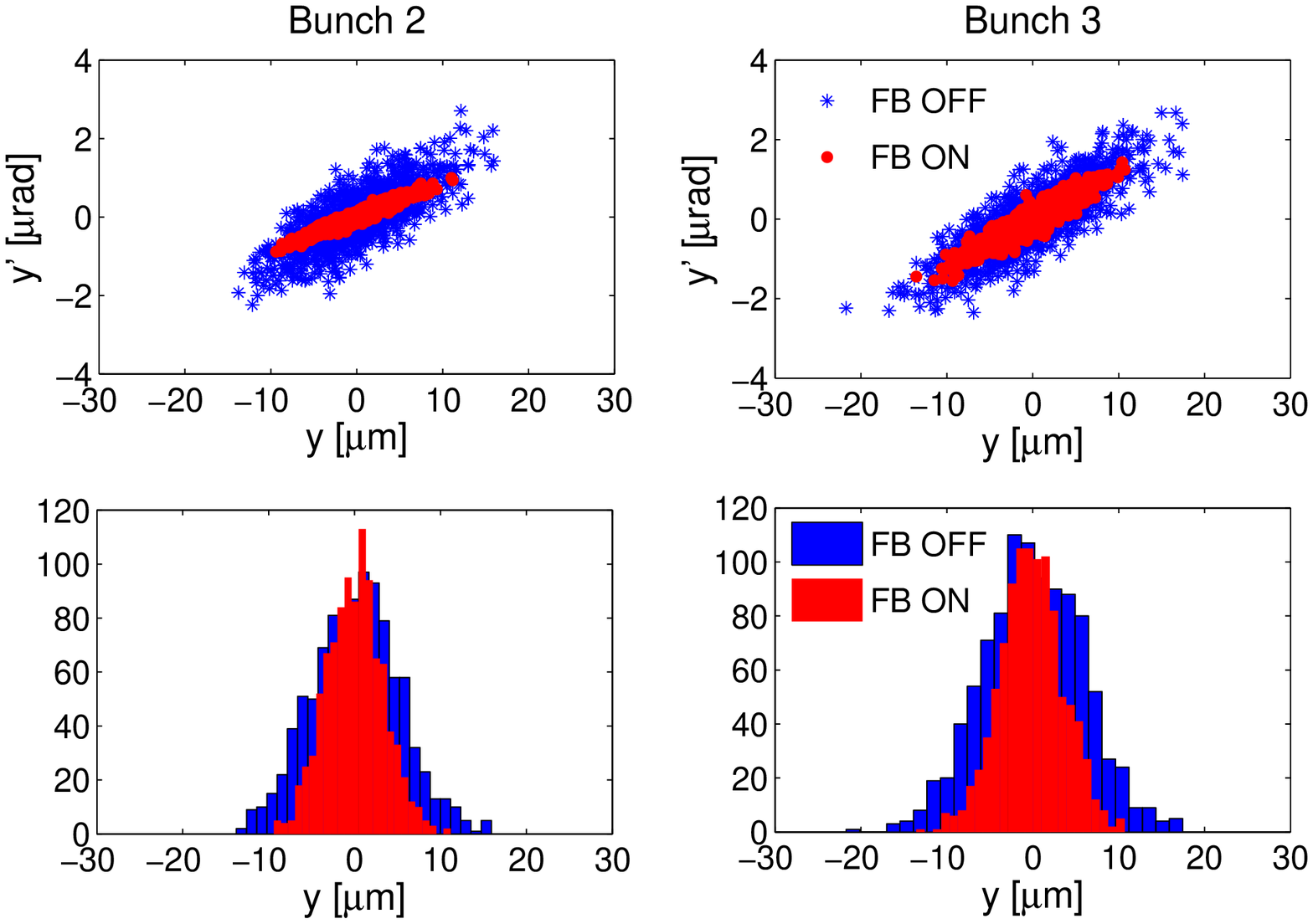}
   \caption{$y$-$y'$ scatter plots (Top) and $y$ distribution (Bottom) at the entrance of the ATF2 FFS from beam tracking simulations for bunch 2 and 3 for 1000 pulses. Performance with FB OFF and ON are compared.}
   \label{OffsetatentranceFFS}
\end{figure}

Figure~\ref{OffsetatentranceFFS} compares the vertical offset distribution at the entrance of the ATF2 Final Focus System (FFS) for bunch 2 and 3 with FONT FB OFF and ON, resulting from the tracking of a distribution of 1000 events from P2. The jitter predictions for bunch 2 and 3 at the entrance of the FFS are summarised in Table~\ref{jitteratentranceFFS}. For instance, with FB ON the position jitter is reduced by approximately a factor 1.5 and the angle jitter by a factor 2 for bunch 2 at the entrance of the FFS. 

\begin{table}[hbt]
   \begin{center}
   \caption{Prediction of the vertical position and angular jitter at the FFS entrance.}
   \begin{tabular}{lcc}
       \hline \hline
       Bunch $\#$ & $\sigma_{\rm FFS}$ [$\mu$m]& $\sigma'_{\rm FFS}$ [$\mu$rad]  \\         
       \hline
        2 (FB OFF/ON) & 5.1/3.3 & 0.7/0.3 \\
        3 (FB OFF/ON) & 5.9/3.7 & 0.9/0.5 \\      
       \hline
   \end{tabular}
   \label{jitteratentranceFFS}
\end{center}
\end{table}

\begin{figure}[htb]
   \centering
   \includegraphics*[width=14cm]{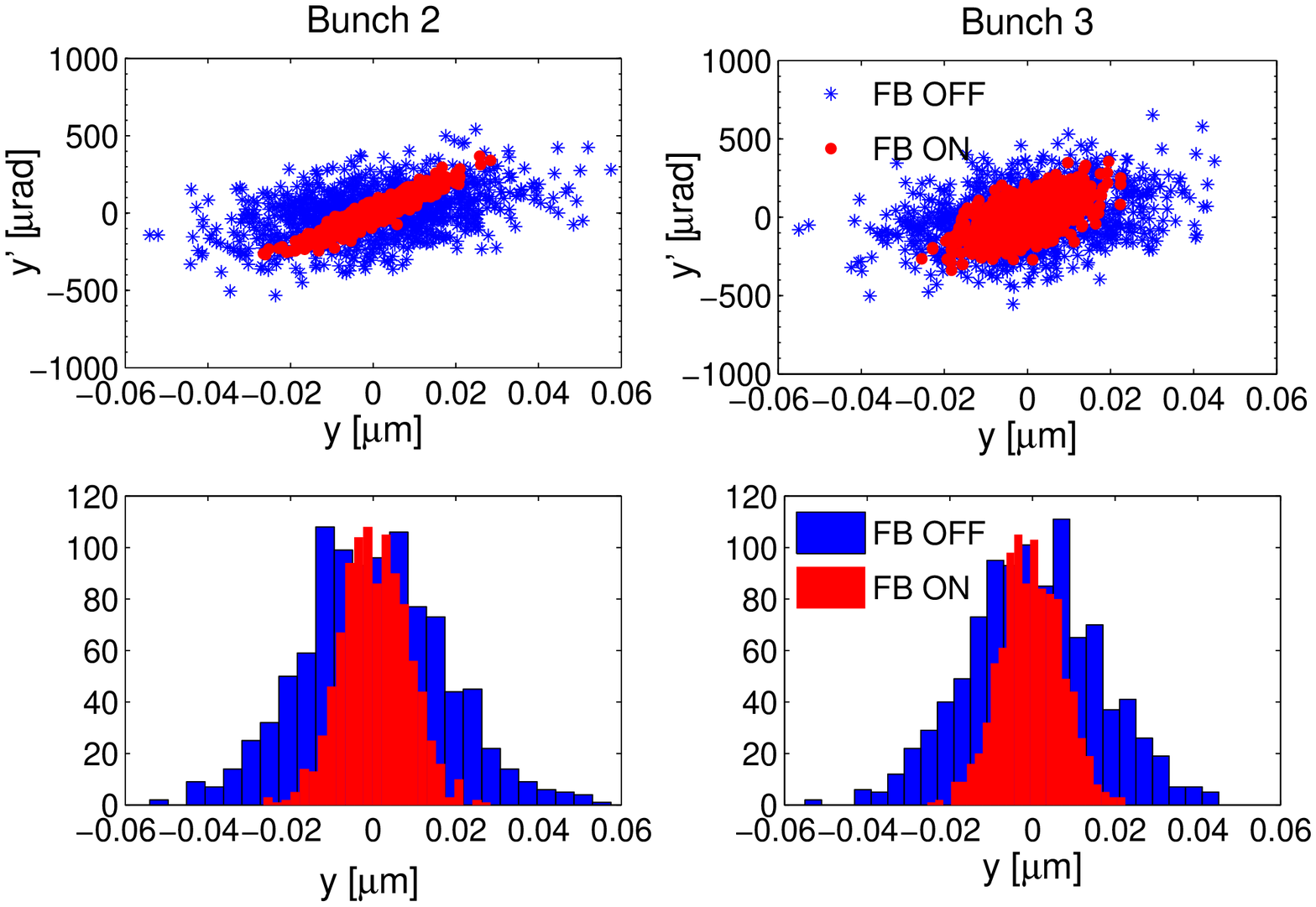}
   \caption{$y$-$y'$ scatter plots (Top) and $y$ distribution (Bottom) at the ATF2 IP from beam tracking simulations for bunch 2 and 3 for 1000 pulses. Performance with FB OFF and ON are compared.}
   \label{OffsetatIP}
\end{figure}

Figure~\ref{OffsetatIP} compares the vertical offset distribution at the IP for bunch 2 and 3 with FONT FB OFF and ON, for 1000 tracked events. The corresponding jitter predictions for bunch 2 and 3 are summarised in Table~\ref{jitteratIP}. With FB ON the position jitter and angle jitter are reduced by approximately a factor 2 and a factor 1.6, respectively, at the IP.    

\begin{table}[hbt]
   \begin{center}
   \caption{Prediction of the vertical position and angular jitter at the IP.}
   \begin{tabular}{lcc}
       \hline \hline
       Bunch $\#$ & $\sigma_{\rm IP}$ [nm]& $\sigma'_{\rm IP}$ [$\mu$rad] \\         
       \hline
        2 (FB OFF/ON) & 17.3/8.0 & 155.6/100.0 \\
        3 (FB OFF/ON) & 16.0/7.5 & 173.7/110.5 \\      
       \hline
   \end{tabular}
   \label{jitteratIP}
\end{center}
\end{table}

\section{Conclusions and Future Plans}
An intra-train FB system has been tested at the ATF2 beam test facility with short ILC-like trains in 3-bunch mode with 154~ns bunch separation. This FB system is placed in the ATF2 extraction line (upstream of the FFS) and corrects the incoming $y$ and $y'$ beam jitter. The FB system performs as expected, reaching a factor 5 position jitter reduction and a factor 3 angle jitter reduction at BPM P2. Simulation studies of jitter propagation have shown that the position and angle jitter are reduced downstream of the FB system. A FB OFF/ON correction ratio of 2 for position jitter and of 1.6 for angle jitter at the ATF2 virtual IP have been predicted by tracking simulations with the nominal ATF2 optics. Results show that the intra-train FB system in the extraction of ATF2 has the potential to stabilise the beam to below 10~nm at the IP. These results are very encouraging and provide an important step towards the achievement of the ATF2 second goal.

Future plans include the following items:

\begin{itemize}

\item Improvement of the FONT BPM resolution for BPM P3 in order to improve the FB correction. The objective is to obtain a resolution $< 1 \mu$m in all three FONT BPMs (P1, P2 and P3). 

\item Pulse-to-pulse jitter measurements (in multi-bunch operation) with other available BPMs downstream of the FB system and comparison with simulation results. 
\item Measurements of the transverse emittance with multi Optical Transition Radiation (multi-OTR) system \cite{OTR} and, simultaneously, measurements of $x$ and $y$ beam jitter with the FONT BPMs. In this way we could get a complete characterisation of the beam in the extraction line, and to investigate possible scaling factors emittance/jitter. 

\item Detailed study of possible incoming pulse-to-pulse and bunch-to-bunch orbit jitter sources in the ATF damping ring or during extraction from the ATF damping ring.
\end{itemize}  

Javier Resta has been supported by grant FPA2010-21456-C02-01 from Ministerio de Ciencia e Innovaci\'on, Spain.


\end{document}